# Multi-Convergence-Angle Ptychography with Simultaneous Strong Contrast and High Resolution

Wei Mao[1,2,#], Weiyang Zhang[1,#], Chen Huang[3,#], Liqi Zhou[1,2,5], Judy S. Kim[3,4], Si Gao[1,6], Yu Lei[2], Xiaopeng Wu[2], Yiming Hu[1], Xudong Pei[1], Weina Fang[7], Xiaoguo Liu[8], Jingdong Song[9], Chunhai Fan[8], Yuefeng Nie[1], Angus I. Kirkland[3,4], Peng Wang[1,2,†]

[1]*National Laboratory of Solid-State Microstructures, Jiangsu Key Laboratory of Artificial Functional Materials, College of Engineering and Applied Sciences and Collaborative Innovation Centre of Advanced Microstructures, Nanjing University, Nanjing 210093, China.*

[2]*Department of Physics, University of Warwick, Coventry CV4 7AL, UK.*

[3]*The Rosalind Franklin Institute, Harwell Campus, Didcot OX11 0QX, UK.*

[4]*Department of Materials, University of Oxford, Parks Road, Oxford OX1 3PH, UK.*

[5]*Institute of Materiobiology, College of Science, Shanghai University, Shanghai, 200444, China.*

[6]*College of Materials Science and Engineering, Nanjing Tech University, Nanjing 210009, China.*

[7]*Shanghai Key Laboratory of Green Chemistry and Chemical Processes, Department of Chemistry, School of Chemistry and Molecular Engineering, East China Normal University, Shanghai 200241, China.*

[8]*School of Chemistry and Chemical Engineering, Frontiers Science Center for Transformative Molecules and National Center for Translational Medicine, Shanghai Jiao Tong University, Shanghai 200240, China.*

[9]*State Key Laboratory of Infectious Disease Prevention and Control, National Institute for Viral Disease Control and Prevention, Chinese Center for Disease Control and Prevention, Beijing 100052, China.*

[#] *These authors contributed equally to this work.*
[†] *Correspondence and requests for materials should be addressed to: Peng Wang (peng.wang.3@warwick.ac.uk).*

**Abstract**

**Advances in bioimaging methods and hardware facilities have revolutionised the determination of numerous biological structures at atomic or near-atomic resolution. Among these developments, electron ptychography has recently attracted considerable attention because of its superior resolution, remarkable sensitivity to light elements, and high electron dose efficiency. Here, we introduce an innovative approach called multi-convergence-angle (MCA) ptychography, which can simultaneously enhance both contrast and resolution with continuous information transfer across a wide spectrum of spatial frequency. Our work provides feasibility of future applications of MCA-ptychography in providing high-quality two-dimensional images as input to three-dimensional reconstruction methods, thereby facilitating more accurate determination of biological structures.**

**Introduction**

Structural biology, which primarily concentrates on the tertiary structure of biological molecules, has played a pivotal role in disease research, drug development, and the investigation of fundamental biological processes. The ongoing technical developments, particularly cryogenic transmission electron microscopy (cryo-TEM), direct electron detectors and three-dimensional (3D) reconstruction algorithms such as single particle analysis (SPA)[1-4] and electron tomography (ET)[5-8], have greatly advanced the structural biology field. However, there is a growing need for more precise determination of the pristine three-dimensional structure of biological molecules using SPA or ET, necessitating high-quality two-dimensional projection image data with clear visibility of both macroscopic morphology and microscopic details. Nonetheless, due to their composition predominantly consisting of light elements, biological protein samples are essentially pure phase objects and are extremely radiation sensitive. Consequently, the recorded images commonly show weak contrast and a low signal-to-noise ratio (SNR). To counteract this, staining[9] is one method used in the preparation of biological samples, although it should be noted that the heavy metal ions and dessication process damage the fine ultrastructure. Alternatively, phase plates such as Zernike[10] and Volta[11] phase plates can be utilized to improve contrast of unstained biological samples[12-14] by introducing a phase shift between scattered and transmitted waves. However, these phase plates suffer from poor reliability, short working lifetime and inconsistent fabrication, and most importantly, the contrast of the phase plates themselves may diminish signal from biological samples and reduce resolution. Another method for improving contrast at low spatial frequencies is using high defocus values[15,16], but this method compromises information transfer at intermediate and high spatial frequencies and ultimately leads to lower resolution[17]. Therefore, there is currently no method for simultaneously obtaining two-dimensional images with both strong contrast and high resolution to satisfy the needs for more precise determination of biological structures.

Ptychography, as initially proposed by Hoppe in 1960s[18], is a phase contrast method related to coherent diffraction imaging. This approach collects a series of far-field diffraction patterns as a function of probe position by scanning a finite-sized probe across the extended sample and subsequently recover the object function using iterative or direct algorithms without any prior knowledge of the probe function. Ptychography has been extensively implemented in light and X-ray optics, and ptychography has also gained considerable interest in electron optics due to its superior resolution[19-22], high contrast light elements detection[23-26], high electron dose efficiency[27-30] and capability of three-dimensional imaging[31-34], which have been recently reviewed[35]. By exploiting these advantages, electron ptychography has been employed for experimentally characterizing biological samples with improved contrast[27,33,36]. However, the contrast transfer function (CTF) of electron ptychography is highly contingent upon the convergence angle of the electron probe[27,36,37]. When characterizing biological samples, a small convergence angle has been used[27] to enhance contrast at the expense of resolution. Hence, conventional electron ptychography has not yet been developed enough to simultaneously achieve both strong contrast and high resolution.

In this work, we propose a novel ptychographic method and experimental setup that enables the incorporation of multiple convergence angle (MCA) data in a single reconstruction. In comparison to conventional, single convergence angle electron ptychography, our ptychographic phase reconstructions demonstrate superior image quality with both strong contrast and high resolution. We further demonstrate this method on a range of samples, from strong scattering and beam-robust materials (e.g., gold particles) at room temperature to weak scattering and beam-sensitive biological materials. Our results show that both contrast and resolution under low-dose conditions can be simultaneously achieved in a single reconstruction. These findings highlight the potential of MCA-ptychography in providing high-quality images, enabling more precise determination of biological structures and the possibility of use in 3D methods.

## Results and Discussions

**MCA-Ptychography Reconstruction Workflow**

The schematic experimental setup of MCA-ptychography is illustrated in Fig. 1a. In contrast to conventional electron ptychography, which employs a single convergence angle (hereafter referred to as single-convergence-angle ptychography or SCA-ptychography for comparison), MCA-ptychography utilizes multiple condenser apertures with varying diameters to modify the convergence angles of the electron probe during the scanning process. A fast direct electron detector synchronized with the scanning system of the electron microscope is used to collect the ptychographic datasets under different convergence angles on the far-field plane, as illustrated in Fig. 1b. However, such a device with the aperture-switching capability is not currently available for electron microscopes. Therefore, as a proof of concept, we performed MCA-ptychography in series by acquiring the first dataset with one aperture for the entire scanning area and subsequently switching to the other aperture to acquire the second dataset for the same area. The two SCA-ptychographic datasets are mixed before reconstructing the object wavefunction via MCA-ptychography reconstruction algorithm, as shown in Fig. 1c (see Supplementary Fig. S11 and Supplementary Note 8 for more details about the MCA-ptychography reconstruction workflow).

**Contrast and Resolution Enhancement**

To experimentally evaluate the contrast and resolution enhancement capability of MCA-ptychography with two different convergence angles in comparison to SCA-ptychography with either of the angles, a standard 50-nm-thick sample of gold particles deposited on a carbon film was employed. The reconstructed phase results obtained through different algorithms were compared. It should be noted that both contrast and resolution are influenced by the electron dose applied. Therefore, for a fair comparison, the beam current and dwell time were adjusted before collecting the SCA-ptychographic data to ensure that the total electron doses in the two datasets were approximately equal. The SCA-ptychographic datasets were divided into two halves in the scheme as shown in Supplementary Fig. S13 and these halves were mixed for MCA-ptychography to ensure that its total electron dose was identical to that of SCA-ptychography. Furthermore, due to the stability of gold particles even under high electron doses, the electron dose can be increased without concern about damaging the sample. This enables a comparison of achievable resolutions using different approaches on the same object without being electron irradiation limited. A series of two SCA-ptychographic datasets with convergent semi-angles, $\alpha$, of 2.25 mrad and 22.50 mrad were acquired from one sample area at 300 kV using a double aberration-corrected FEI Titan Cubed 60-300 STEM (This microscope was used in all experiments unless specified otherwise). Ptychographic datasets were recorded on a 1024 × 1024 pixel Gatan Oneview Camera (Further experimental settings in Supplementary Note 2).

SCA-ptychographic reconstructed phases for $\alpha$ of 2.25 mrad and 22.50 mrad are presented in Fig. 2a and b, respectively. Magnified views of subregions indicated by the yellow squares in Fig. 2a and b, are shown in Fig. 2d and e, along with the power spectra calculated from Fig. 2a and b displayed in Fig. 2g and h, respectively. In comparison, the reconstructed phase at 2.25 mrad captures the overall shape of the gold particles with stronger contrast compared to the reconstruction at 22.5 mrad. However, despite weaker contrast, the reconstructed phase at 22.50 mrad achieves higher resolution as expected for focused probe-based data collection. The power spectrum in Fig. 2h reveals the presence of {311} reflections, corresponding to a real-space resolution of 0.93 Å, whereas no reflections are observed in the 2.25 mrad generated power spectrum in Fig. 2g. This observation is further supported by line profiles extracted across the gold particles, as shown in Fig. 2j. Specifically, the curve derived from the phase result at 2.25 mrad exhibits strong but low-frequency fluctuations, representing the overall shape of gold nanoparticles. In contrast, the curve obtained from the reconstructed phase at 22.50 mrad displays relatively weak phase shifts but possesses periodic high-frequency oscillations.

Following the procedure illustrated in Fig. 1, MCA-ptychographic reconstruction was employed and the reconstructed phase is shown in Fig. 2c. Magnified views and corresponding power spectra are presented in Fig. 2f and Fig. 2i, respectively. Notably, the MCA-ptychographic phase result not only captures the overall shape of the gold particles with strong contrast but also resolves their lattice fringes. This is evidenced by the presence of {400} reflections in the power spectra shown in Fig. 2i, indicating a real-space resolution of 1.02 Å. The mixed-angle line profile displayed in Fig. 2j, exhibiting strong low-frequency phase fluctuations along with periodic high-frequency oscillations, further demonstrates the capability of MCA-ptychography to enhance contrast and resolution simultaneously. It provides comprehensive structural information ranging from overall morphology to microscopic details.

To investigate the impact of different reconstruction algorithms on information transfer, numerically calculated contrast transfer functions (CTFs) are shown in Fig. 2k. Further details of CTF calculation are given in Supplementary Note 6. The CTF of SCA-ptychography at $\alpha$ of 2.25 mrad (blue curve) exhibits strong low spatial frequency transfer; however, its cut-off frequency is low, indicating that the reconstructed phase at 2.25 mrad possesses strong contrast but limited resolution. In contrast, the CTF of SCA-ptychography at $\alpha$ of 22.50 mrad (orange curve) has a larger cut-off frequency, despite weak low spatial frequency information transfer, suggesting that the reconstructed phase at 22.50 mrad obtains high resolution but weak contrast. This observation is consistent with our previous work[27]. Notably, the CTF of MCA-ptychography (purple curve) not only shows strong low spatial frequency transfer, but also extends its cut-off frequency, highlighting the ability of MCA-ptychography to simultaneously achieve strong contrast and high resolution.

It is worth mentioning that there is a decrease in the signal intensity at high spatial frequencies in the MCA-ptychographic CTF compared to the SCA-ptychographic CTF at 22.50 mrad, which can be attributed to the fact that only half of the dataset at 22.50 mrad was used in MCA-ptychography. Consequently, the SNR at high spatial frequencies in MCA-ptychography is attenuated, resulting in slightly lower resolution when compared to SCA-ptychography at 22.50 mrad - from 0.93 Å to 1.02 Å.

**Room Temperature Biological Imaging**

Subsequently, an investigation into the efficiency of MCA-ptychography in characterizing biological specimens was initiated. To be generally applicable to structural studies of biological materials, especially cellular ultrastructures, MCA-ptychography must be capable of recovering information from a large field of view (FOV). Therefore, 30-nm-thick ultra-thin sections of unstained resin-embedded human embryo kidney cells 293 (HEK 293) infected with Adenovirus type 5 (Ad5) were studied. Further details about the sample preparation procedure can be found in Supplementary Note 1, which are also described in our previous work[27]. Ptychographic datasets at $\alpha$ of 1.18 mrad and 11.80 mrad were collected at 60 kV and at a dose of approximate 350 e$^-$/Å$^2$. Additionally, scanning step size was increased to achieve a FOV of 1.02 um, reaching the micrometer scale (Details of the relevant experimental settings are provided in Supplementary Note 3).

The comparison between the reconstructed phase obtained from SCA-ptychographic datasets at 1.18 mrad, 11.80 mrad and MCA-ptychographic dataset is presented in Fig. 3. The SCA-ptychographic reconstructed phase at 1.18 mrad (Fig. 3a) shows clear visibility of free ribosomes and Ad5 viral particles with DNA contained inside the icosahedral capsid, while these ultrastructural features are hardly recognizable in the SCA-ptychographic reconstructed phase at 11.80 mrad (Fig. 3b). Conversely, the Fourier ring correlation (FRC) using a 2σ criteria[38,39], as illustrated in Fig. 3d, demonstrates that the reconstructed phase at 11.80 mrad achieved a higher resolution of 0.96 nm than lower semi-angle data. The MCA-ptychographic reconstructed phase, as depicted in Fig. 3c, not only presents the overall morphology of the free ribosomes and Ad5 viral particles with strong contrast, but also achieves high resolution of 1.45 nm.

It should be noted that the splitting of dataset in FRC analysis reduces the probe overlap ratio and total dose in the resulting sub-datasets, thereby leading to a conservative estimation of spatial resolution. For comparison, conventional defocused TEM images of the same sample area were captured at 60 kV within a range of defocus value from -0.2 μm to -10.4 μm, as shown in Supplementary Fig. S4. Due to the high defocus values, rapid oscillations were observed in the diffractograms, resulting in contrast reversals at intermediate and high spatial frequencies. Consequently, an optimal resolution of 3.42 nm was achieved at a defocus of -0.2 μm in conventional TEM, despite the weak contrast exhibited by Ad5 viral particles.

The nanomaterials have been widely used in desired cell responses, like targeting, and drug delivery. Understanding know how inorganic nanomaterials interact with the organic components of cells at the atomic level, including their orientation and facets, is crucial for accurately controlling cell behavior by tuning of nanomaterial properties for bioactivity, biocompatibility, and safety[40-42]. To address this need, an imaging technique is required that can simultaneously visualize both the biological and non-biological components in their native state without staining[33]. This technique should have the capability to resolve the crystallography of non-biological components at the atomic resolution. To demonstrate the capabilities of MCA-ptychography, a sample of unstained DNA origami with gold particles suspended on an ultrathin carbon film at room temperature was used. Ptychographic datasets at 2.25 mrad and 22.50 mrad at 60kV (See Supplementary Note 4 for detailed experimental settings).

The reconstructed phase results from different datasets are presented in Fig. 4. The SCA-ptychographic reconstructed phase at 2.25 mrad reveals clear overall morphology of both DNA strands and gold nanoparticles (Fig. 4a). Furthermore, a layer of DNA encapsulating the gold nanoparticles is resolved with directly interpretable contrast (Fig. 4d), indicating high information transfer at low frequencies. In contrast, the reconstructed phase at 22.50 mrad (Fig. 4b) only exhibits the edge of the gold nanoparticles but fails to directly resolve the DNA strands, as evidenced by line profiles (Fig. 4p) extracted across the DNA stands in Fig. 4a and b, respectively. This indicates that the reconstructed phase at the higher angle of 22.50 mrad lacks low frequency transfer. However, for high frequency information, it transfers strongly. The magnified image of an Au particle (Fig. 4f), extracted from Fig. 4b, and its power spectrum (Fig. 4g) exhibit lattice fringes and additional reflections from (020) planes with a lattice spacing of 2.04 Å, respectively, in comparison to those of reconstructed phase at 2.25 mrad (Fig. 4 d and e). Detailed crystallography information on nanoparticles can be obtained such that the Au particle is a single crystal with an orientation of $[\bar{1}01]$. Therefore, using SCA-ptychography only provides partial information on the sample, suffering from difficulties in completed structural analysis for similar types of materials system.

Beneficially, MCA-ptychography offers the information transfer over a much wider frequency range. As shown in its reconstructed phase (Fig. 4c and h), not only are gold nanoparticles and DNA strands are resolved, but lattice fringes and crystal orientation of the nanoparticles are also identified (Fig. 4i). The results from MCA-ptychography possess information with a wider bandwidth of spatial frequencies covering the total bandwidth combined from two SCA-ptychography sets, band-pass filtering was implemented to their phases by applying a mask from 0.0-0.6 nm$^{-1}$ to their power spectra (Fig. 4e, g and i). By inversely Fourier transforming the masked or unmarked region of the power spectra, three pairs of low-pass (Fig. 4j, l and n) and high-pass (Fig. 4k, m and o) images were obtained, respectively. Comparing those images, only the pair filtered from MCA-ptychographic reconstruction has the bandwidth low enough to reveal the overall shape of the Au nanoparticle (Fig. 4n) and high enough to resolve the lattice fringes (Fig. 4o). This observation is easily visualised by line profiles extracted across the Au nanoparticle (Fig. 4q). Therefore, this MCA-ptychography holds great potential to reveal not only high-contrast 3D morphological framework of organic-inorganic hybrid-structures[33], but also to provide correlative crystallographic information about inorganic components at an atomic scale, which can offer a deep understanding how inorganic nanomaterials interact with the organic components of cells at the atomic level.

**Low Dose Simulations**

To investigate the impact of electron dose on the quality of MCA-ptychographic reconstructed results, dose-dependent multislice simulations[43,44] were conducted using a model of a GroEL protein (PDB: 2eu1) suspended on a monolayer graphene substrate. Supplementary Note 5 provides additional information about the Multislice simulation procedures, and corresponding atomic potential models displayed in Supplementary Fig. S8. The ptychographic datasets were simulated with convergence semi-angles of 3 and 15 mrad, respectively. Fig. 5 presents a comparison of reconstructed phases from SCA-ptychography datasets with convergence semi-angles of 3 mrad (Left Column), 15 mrad (Middle Column), and the MCA-ptychography dataset from 3 mrad and 15 mrad (Right Column) as a function of electron dose. At a fluence of 38 e$^-$/Å$^2$ (upper row), the MCA-ptychography reconstruction (Fig. 5e) exhibits both low spatial frequency information from the overall shape of the GroEL protein and high spatial frequency information from the atomic structures of the protein and the graphene beneath. As the fluence decreases to 11 e$^-$/Å$^2$, both the visibility of graphene lattices and atomic structures of the protein fade (Fig. 5f-h). The observed degradation of high frequency information at low doses is consistent with our previous experimental findings on 2D MoS$_2$[28]. However, the contrast of the overall shape of the GroEL protein is maintained when using small convergence angles (Fig. 5f), whereas the overall shape of the GroEL protein becomes difficult to discern when using large convergence angles (Fig. 5g). Using MCA-ptychography, the reconstructed phase results can still possess both clear morphological contrast of protein and lattice information of monolayer graphene (Fig. 5h), as evidenced by the power spectra and the line profiles provided in Supplementary Fig. S10, demonstrating that MCA-ptychography is robust under low dose conditions.

## Conclusions

In conclusion, we propose and demonstrate the effectiveness of a novel ptychographic method, MCA-ptychography, for achieving high-quality two-dimensional images with both strong contrast and high resolution simultaneously across a broad spatial frequency spectrum. The experimental results on various samples, including biological specimens and organic-inorganic hybridstructures, highlight the potential of MCA-ptychography in providing more precise determination of biological structures. The method's capability to incorporate multiple convergence angles allows it to overcome the limitations of conventional electron ptychography, making it a valuable tool for advancing structural biology and enabling simultaneous investigations of morphology and atomic structure. We anticipate further breakthroughs in this field, particularly in integrating MCA-ptychography with low-contrast unstained cryo-electron microscopy (cryo-EM) specimens. This integration aims to provide improved 2D projection images characterized by enhanced contrast and resolution, thereby benefiting subsequent 3D-reconstruction methods. This novel approach holds the potential to elevate the precision of 3D reconstructions for various biological objects, including viruses, proteins, and cells.

## Methods

**Samples.** Three samples were used: the gold nanoparticles used in Fig. 2 are the commercial gold grating replica specimen bought from TED PELLA, INC. Full details about the preparation procedure of ultra-thin sections of Adenovirus-infected cells (Fig. 3) can be found in Supplementary Note 1. DNA origami with a tetrahedral structure with gold nanoparticles inlayed (Fig. 4) was prepared using one-step self-assembly method[45]. DNA origami was unstained and suspended on a 50-mesh ultra-thin carbon film EM grid. The EM grid was rinsed three times using the deionized water to remove salt crystals from the buffer solution and air-dried in room temperature.

**Ptychographic data acquisition.** All the ptychographic datasets under different convergence semi-angles were collected in a scanning diffraction mode on a double aberration-corrected FEI Titan Cubed 60-300. Specifically, ptychographic datasets of gold nanoparticles were collected at 300 kV with data recorded on a 1024 × 1024 Gatan Oneview Camera. Ptychographic datasets of Adenovirus-infected cells were also collected using a 1024 × 1024 Gatan Oneview Camera at 60 kV. The SCA-ptychographic dataset of DNA origami at α of 2.25 mrad was collected at 60 kV using an electron microscope pixel-array detector (EMPAD)[46], while the SCA-ptychographic dataset of DNA origami at 22.50 mrad was recorded on a 1024 × 1024 Gatan Oneview Camera. Detailed additional experimental settings are given in Supplementary Note 2-4.

**Multi-Convergence-Angle (MCA) ptychographic reconstruction.** We adapted the conventional extended ptychographical iterative engine (ePIE) algorithm[47], which employs only one convergence angle, to be applicable for the reconstruction of datasets under multiple convergence angles. The MCA-ptychographic reconstruction procedure is summaried below.

Assuming that we have a total of *n* distinct convergence angles, we define probe functions under these convergence angles, $\{P_1, P_2, \cdots, P_n\}$, an object function, *O*, object exit wave functions, $\{\Phi_1, \Phi_2, \cdots, \Phi_n\}$, diffraction patterns, $\{\Psi_1, \Psi_2, \cdots, \Psi_n\}$, the measured intensities of these diffraction patterns, $\{I_1, I_2, \cdots, I_n\}$, coordinates in real space, $\vec{r}$, and in reciprocal space, $\vec{k}$, and the probe position, $\vec{r_p}$.

At a probe postion, estimates of the object exit wave functions under different convergence angles are formed as follows:

$$\Phi_{i,j}(\vec{r}) = P_{i,j}(\vec{r} - \vec{r_p})O_j(\vec{r}) \quad (i = 1, 2, \cdots, n) \tag{1}$$

where the subscript $j$ indicates the j-th iteration.

Initial guesses of the diffraction patterns in the far field are calculated as:

$$\Psi_{i,j}(\vec{k}) = \mathcal{F}(\Phi_{i,j}(\vec{r})) \quad (i = 1, 2, \cdots, n) \tag{2}$$

where the $\mathcal{F}$ denotes the Fourier transform.

After replacing the amplitude with the experimentally measured intensity, updated exit wave functions can be generated as:

$$\Phi'_{i,j}(\vec{r}) = \mathcal{F}^{-1}(\frac{\Psi_{i,j}(\vec{k})}{|\Psi_{i,j}(\vec{k})|}\sqrt{I_i(\vec{k})}) \quad (i = 1, 2, \cdots, n) \tag{3}$$

Finally, an unpdated object function and two updated probe functions are computed as:

$$O_{j+1}(\vec{r}) = O_j(\vec{r}) + \gamma_i \frac{P^*_{i,j}(\vec{r} - \vec{r_p})}{|P_{i,j}(\vec{r} - \vec{r_p})|^2_{max}}(\Phi'_{i,j}(\vec{r}) - \Phi_{i,j}(\vec{r})) \tag{4}$$

$$P_{i,j+1}(\vec{r}) = P_{i,j}(\vec{r}) + \beta_i \frac{O^*_j(\vec{r} + \vec{r_p})}{|O_j(\vec{r} + \vec{r_p})|^2_{max}}(\Phi'_{i,j}(\vec{r}) - \Phi_{i,j}(\vec{r})) \tag{5}$$

$$(i = 1, 2, \cdots, n)$$

The step size of the update is adjusted by the parameters $\{\gamma_1, \gamma_2, \cdots, \gamma_n\}$ and $\{\beta_1, \beta_2, \cdots, \beta_n\}$. In all the experiments presented in this work, $n = 2$. Further details of the reconstruction used in this work are given in Supplementary Note 8.

## Data availability

The data that support this study are available from the corresponding author upon reasonable request.

## Acknowlegements

P.W. acknowledges funding from BBSRC International Institutional Partnership Fund and the University of Warwick Research Development Fund (RDF) 2021-22 Science Development Award. L.Q.Z. acknowledges the funding from the National Natural Science Foundation of China (32301156).

## Author contributions

P.W. conceived the overall project. W.M. and W.Y.Z. developed the MCA-ptychographic algorithm and conducted ptychographic data acquisition at room temperature. W.M., C.H. and P.W. performed the ptychographic experiments at low temperature. L.Q.Z., W.N.F., X.G.L., J.D.S. and C.H.F. provided the samples. W.M. carried out ptychographic simulations. S.G., Y.L., X.P.W., Y.M.H. and X.D.P. assisted with the ptychographic reconstructions and multislice simulations. W.M., W.Y.Z., C.H., L.Q.Z., J.S.K., S.G., Y.F.N., A.I.K. and P.W. discussed the data analysis. W.M., J.S.K. and P.W. wrote and edited the manuscript. All authors discussed the results and commented on the manuscript.

## Competing interests

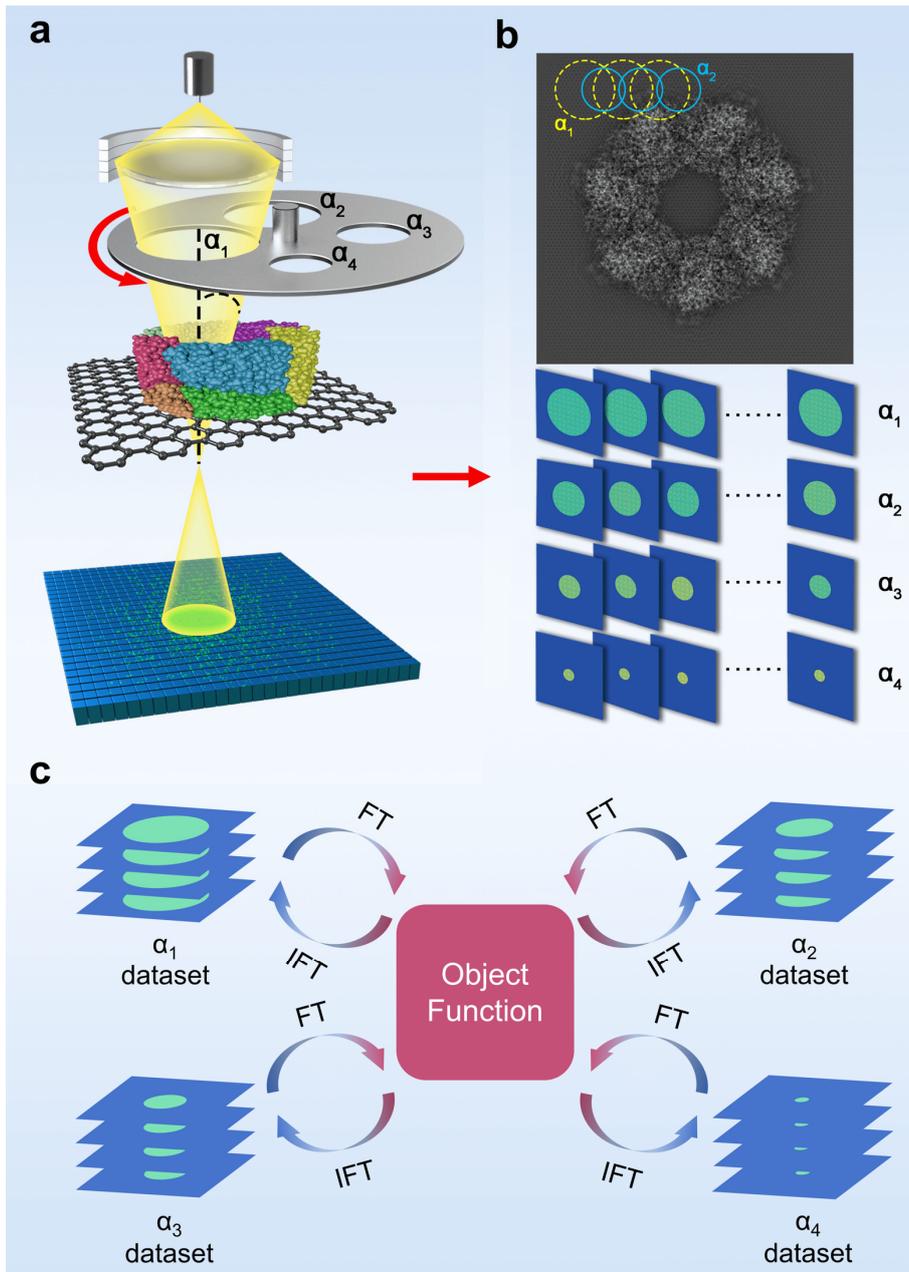

**Fig. 1: Multi-Convergence-Angle (MCA) ptychography experimental setup and workflow. (a)** Schematic diagram of the optical configuration. **(b)** Top view of a simulated GroEL protein suspended on a monolayer graphene substrate. Diffraction datasets under different convergence angles are shown below. **(c)** Schematic diagram of MCA-ptychographic algorithm, wherein the arrows indicate the (inverse) Fourier transforms and desired input-output information.

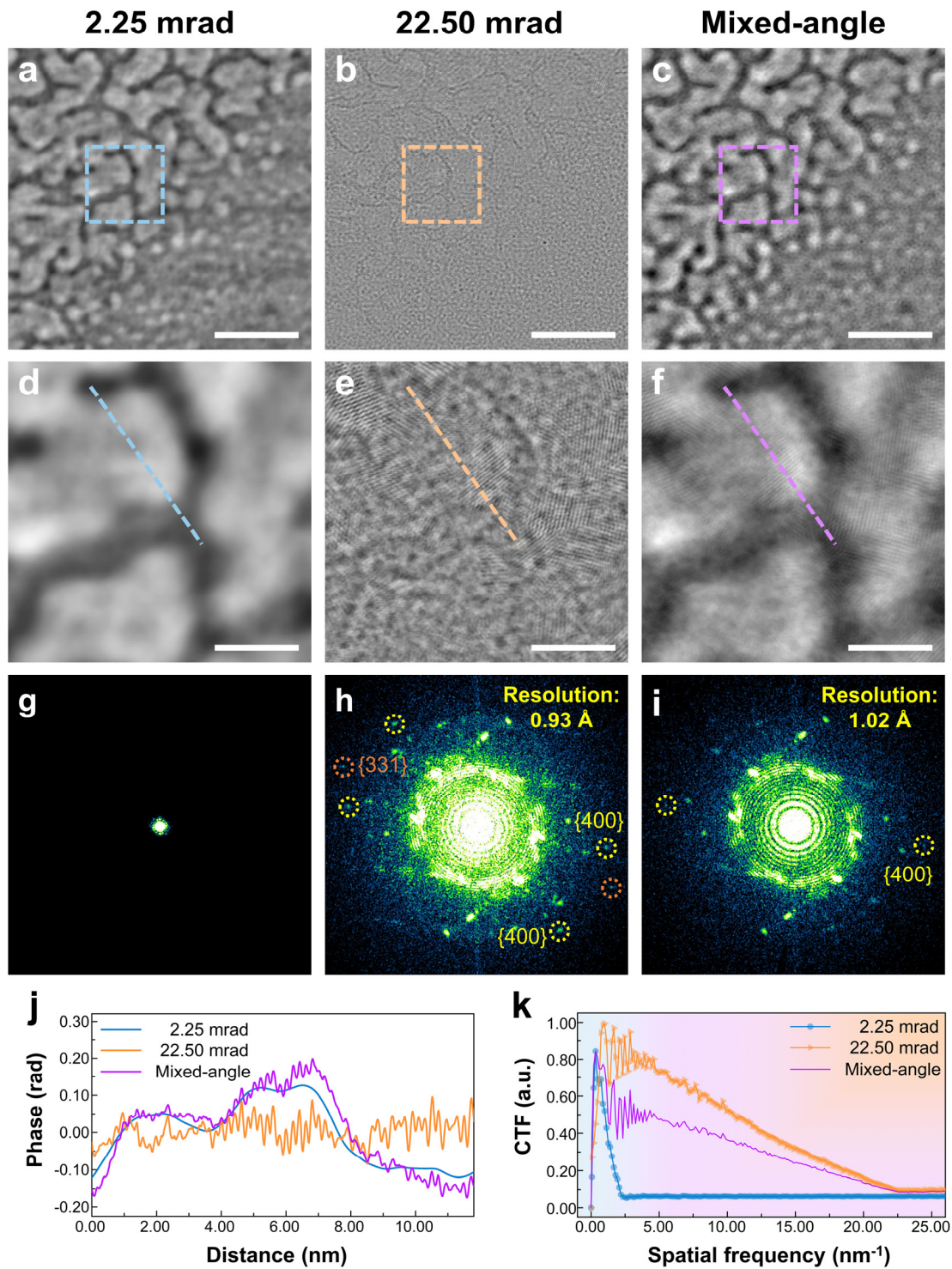

**Fig. 2: Contrast and resolution enhancement evaluation using gold nanoparticles.** Ptychographic reconstructed phase results of gold nanoparticles obtained from datasets under convergence semi-angles of **(a)** 2.25 mrad and **(b)** 22.50 mrad. **(c)** The MCA-ptychographic reconstructed result of gold nanoparticles. **(d-f)** Magnified images of the regions marked with yellow squares in **(a-c)**, respectively. **(g-i)** Power spectra

calculated from **(a-c)**. **(j)** Line profiles extracted from the positions indicated by yellow dashed lines in **(d-f)**. **(k)** Calculated ptychographic CTFs under identical conditions, with Poisson noise added. Scale bars: 20 nm in **(a)** to **(c)**, 5 nm in **(d)** to **(f)**.

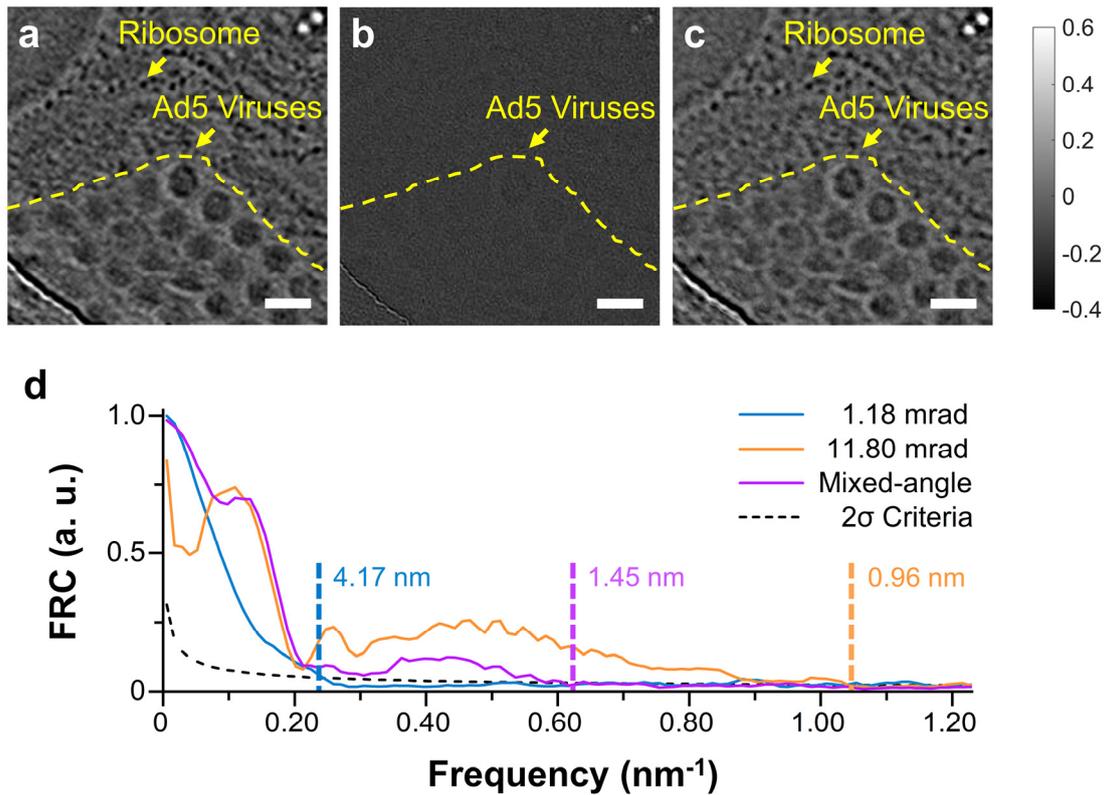

**Fig. 3: Ptychographic reconstructions of Adenovirus type 5 (Ad5) infected cells at room temperature.** SCA-ptychographic reconstructed phases at **(a)** 1.18 mrad and **(b)** 11.80 mrad. **(c)** MCA-ptychographic reconstructed phase result of Ad5 infected cells. **(d)** Resolution estimation of the ptychographic reconstructions using Fourier ring correlation with a 2σ criteria. Scale bars: 150 nm in **(a)** to **(c)**.

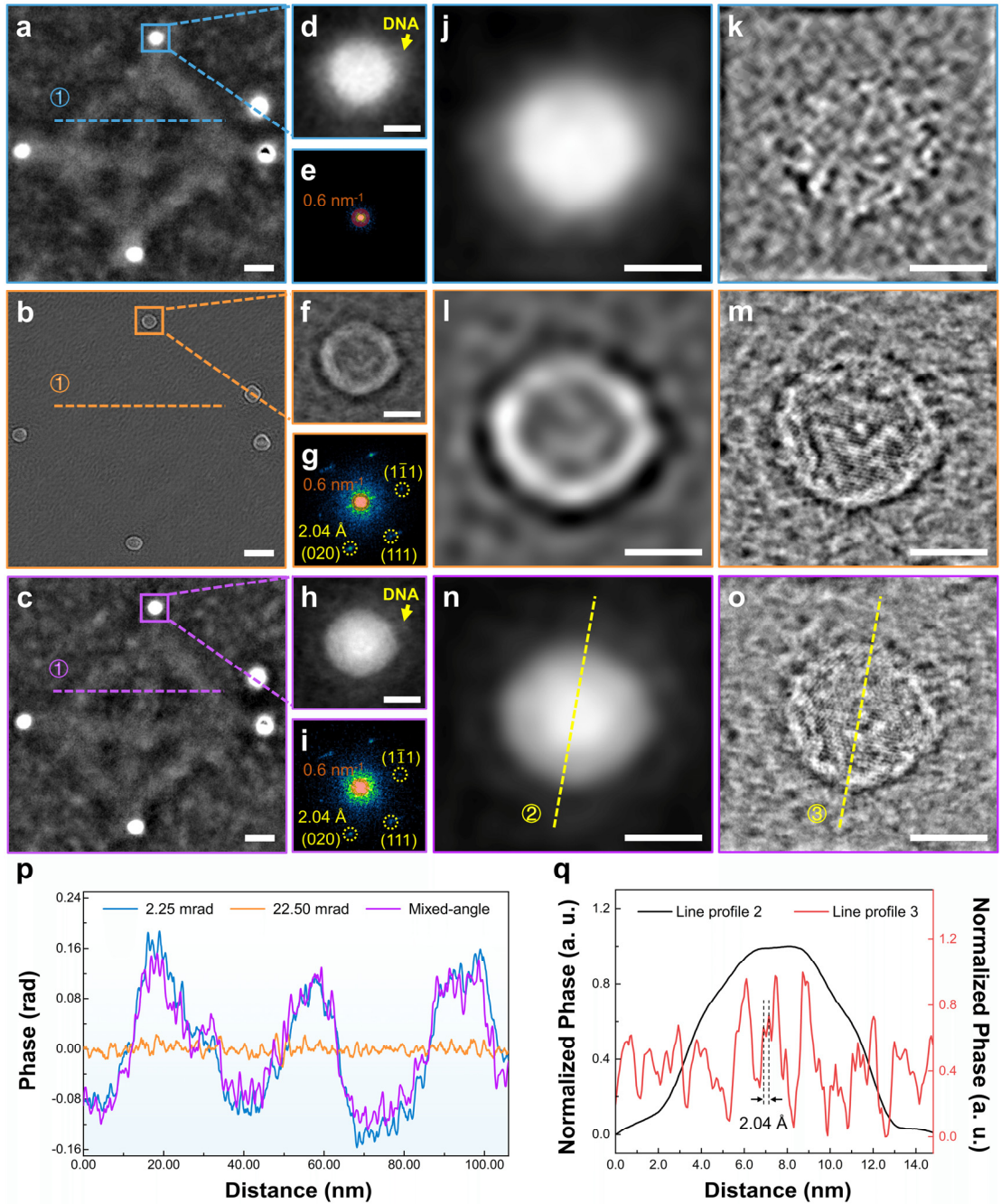

**Fig. 4: Ptychographic reconstructions of DNA origami at room temperature.** SCA-ptychographic reconstructed phase results using datasets under convergence semi-angles of **(a)** 2.25 mrad and **(b)** 22.50 mrad. **(c)** MCA-ptychographic reconstructed phase result of DNA origami. **(d, f, h)** enlarged regions of the phase marked by squares in **(a-c)**, respectively, with corresponding power spectra shown in **(e, g, i)**. Band-pass-filtered reconstructed phases of the gold nanoparticle calculated from **(j, l, n)** 0.0-0.6 nm$^{-1}$ and **(k, m, o)** beyond 0.6 nm$^{-1}$. **(p)** Line profiles across DNA strands, as indicated

by dashed lines in **(a-c)**. **(q)** Line profiles across gold nanoparticles in the MCA-ptychographic result, as indicated by dashed lines in **(n)** and **(o)**. Scale bars: 20 nm in **(a)** to **(c)**, 5 nm in **(d, f, h)** and **(j)** to **(o)**.

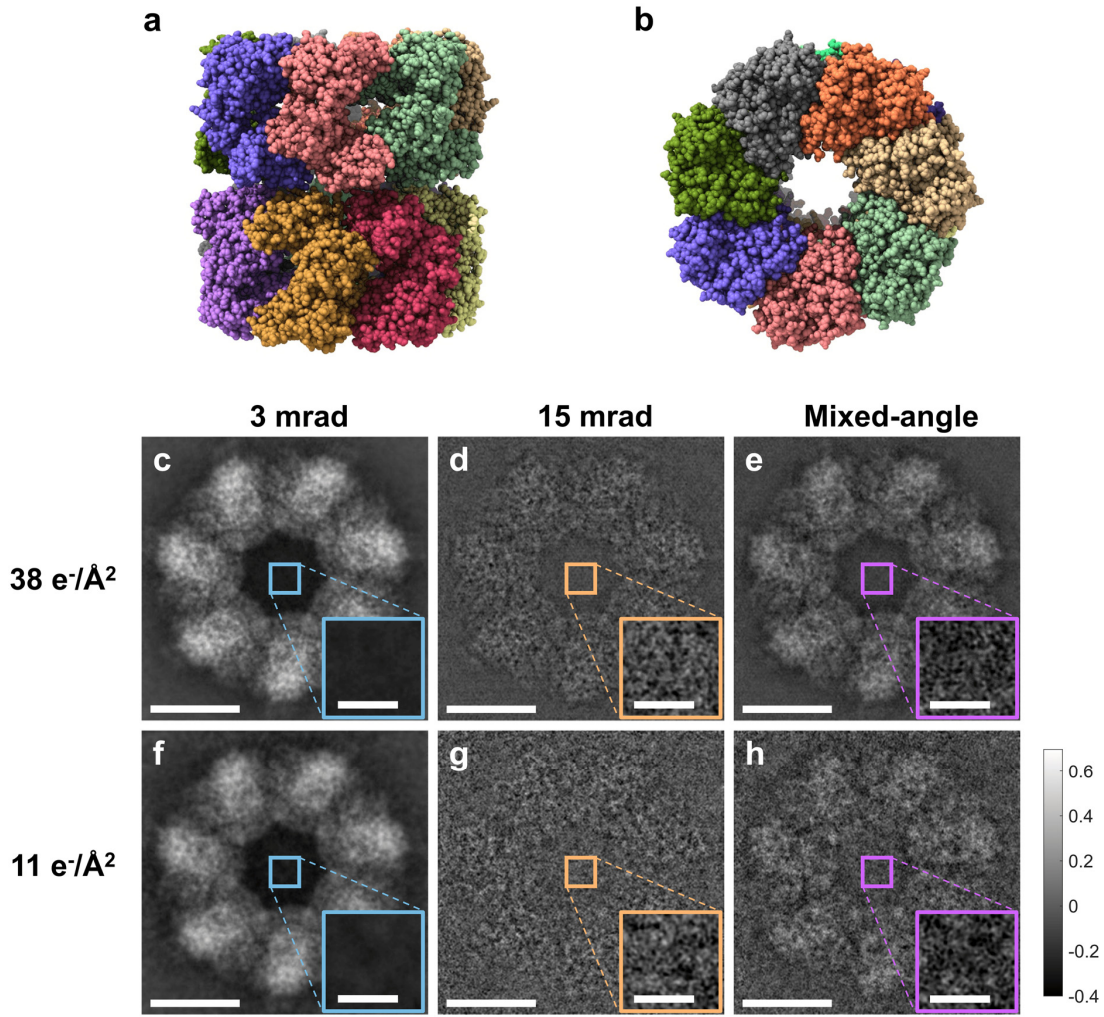

**Fig. 5:** Comparison between SCA-ptychography and MCA-ptychography at low electron doses. **(a)** Side view and **(b)** top view of GroEL protein model (PDB: 2eu1). Ptychographic reconstructions of a simulated GroEL protein suspended on a monolayer graphene under convergence semi-angles of **(c, f)** 3 mrad, **(d, g)** 15 mrad. **(e, h)** present the MCA-ptychographic reconstructed phase results from the combined dataset. The electron dose is decreased from 38 e$^-$/Å$^2$ to 11 e$^-$/Å$^2$. The insets within each figure provide magnified views of the regions marked by squares. Scale bars: 5 nm in **(c)** to **(h)** and 1 nm in insets.